\RequirePackage{fix-cm}

\documentclass[twocolumn,epjc3]{svjour3}
\RequirePackage[T1]{fontenc}
\smartqed
\RequirePackage{graphicx}
\RequirePackage{mathptmx}
\RequirePackage{flushend}
\RequirePackage{latexsym}
\RequirePackage{bm}
\RequirePackage[numbers,sort&compress]{natbib}
\RequirePackage[colorlinks,citecolor=blue,urlcolor=blue,linkcolor=blue]{hyperref}

\journalname{Eur. Phys. J. C}

\begin{document}

\title{Rapidly rotating pulsar radiation in vacuum nonlinear electrodynamics}

\author{V.I.Denisov\thanksref{addr1} \and
        I.P.Denisova\thanksref{addr2} \and
        A.B.Pimenov\thanksref{addr1} \and
        V.A.Soklolov\thanksref{e1,addr1}
        }

\thankstext{e1}{sokolov.sev@inbox.ru}

\institute{Physics Department, Moscow State University, Moscow 119991, Russia\label{addr1}
           \and
           Moscow Aviation Institute (National Research University)
           Volokolamskoe Highway 4, Moscow 125993,  Russia\label{addr2}
          }

\date{Received: date / Accepted: date}

\maketitle

\begin{abstract}
In this paper we investigate vacuum nonlinear electrodynamics
corrections on rapidly rotating pulsar radiation and spin-down in
the perturbative QED approach (post-Max\-well\-ian approximation).
An analytical expression for the pulsar's radiation intensity has
been obtained and analyzed.
\end{abstract}

\section{Introduction}\label{Sect1}
Vacuum nonlinear electrodynamics  effects is an object that piques a great interest in contemporary physics
\cite{a1,a2,a3,a4}. First of all it is related to the emerging opportunities of experimental research in
terrestrial conditions using extreme laser facilities like Extreme Light Infrastructure (ELI) \cite{a5,a6,a7},
Helm\-holtz International Beamline for Extreme Fields (HIBEF) \cite{a8}. It opens up new possibilities in
fundamental physics tests \cite{a9,a10,a11} with an extremal electromagnetic field intensities and particle
accelerations that have never been obtained before.

At the same time investigation of vacuum nonlinear electrodynamics effects in astrophysics gives us an
additional opportunity to carry out a versatile research using natural extreme regimes of strong
electromagnetic and gravitational fields with intensities unavailable yet in laboratory conditions. Compact
astrophysical objects with a strong field, such as pulsars and magnetars are best suited for  vacuum nonlinear
electrodynamics researches. Nowadays, there are many
vacuum nonlinear electrodynamics effects predicted
in the pulsar's neighborhood. For example vacuum electron-positron pairs production \cite{a12} and photon
splitting~\cite{a13}, photon frequency doubling \cite{a14}, light by light scattering and vacuum birefringence
\cite{a15}, transient radiation ray bending \cite{a16,a17} and normal waves delay \cite{a18}. Some of
predicted effects are indirectly confirmed by astrophysical observations. For instance the evidence of the
absence of the high-field (surface fields more then $B_p>10^{13}G$) radio loud pulsars can be explained by
pair-production suppression due to photon splitting \cite{a19}.

In this paper we calculate vacuum nonlinear
electrodynamics corrections to electromagnetic radiation of
rapidly rotating pulsar and analyze pulsar
spin-down under these corrections.

This paper is organized as follows. In Sec.~\ref{Sect2},
we present vacuum nonlinear electrodynamics models and discuss
their main physical properties and predictions. In Sec.~\ref{Sect3} pulsar radiation in post-Maxwellian approximation is
calculated. Sec.~\ref{Sect4} is devoted to analysis of
pulsar spin-down under vacuum nonlinear electrodynamics influence.
In the last section we summarize our results.

\section{Vacuum nonlinear electrodynamics theoretical models}\label{Sect2}
Modern theoretical models of nonlinear vacuum electrodynamics suppose that electromagnetic field Lagrange function density $L=L(I_{(2)},I_{(4)})$ depends on both independent invariants $I_{(2)}=F_{ik}F^{ki}$ and
$I_{(4)}=F_{ik}F^{kl}F_{lm}F^{mi}$ of the electromagnetic field tensor $F_{ik}$.
The specific relationship between Lagrange function and the
invariants depends on theoretical model choice. Nowadays the most promising models are Born-Infeld and Heisenberg-Euler electrodynamics.

Born-Infeld electrodynamics  is a phenomenological theory originating from the requirement of self-energy
finiteness for pointlike electrical charge \cite{a21}. In subsequent studies, the attempts of quantization
were performed \cite{a22,a23} and also it was revealed that Born-Infeld theory describes dynamics of
electromagnetic fields on D-branes in string theory \cite{a24,a25,a26}. As the main features of Born-Infeld
electrodynamics one can note the absence of birefringence (however there are modifications of the Born-Infeld
theory  \cite{a27} with the vacuum birefringence predictions) and dichroism for electromagnetic waves
propagating in external electromagnetic field \cite{a28}. Furthermore, this theory has a distinctive feature
-- the value of electric field depends on the direction of approach to the point-like charge. This property
was noted by the authors of the theory and also eliminated by them in the subsequent model
development\cite{a29}.

Lagrangian function in  Born-Infeld electrodynamics has the following form:
\begin{equation}\label{BI_Lagr}
L=-{1\over 4\pi a^2} \Big\{\sqrt{[1-{a^2\over 2} I_{(2)}-{a^4\over 4} I_{(4)}+ {a^4\over 8}
I_{(2)}^2]}-1\Big\} ,\label{eqna}
\end{equation}
where $a$ is a characteristic constant of theory, the inverse value of which has a
meaning of  maximum electric  field for the point-like charge. For this constant only the following estimation
is known: $a^2<1.2\cdot10^{-32}$ G$^{-2}$.

The other nonlinear vacuum electrodynamics -- Hei\-sen\-berg-Euler model \cite{a15, a30} was derived in
quantum field theory and describes one-loop radiative corrections caused by vacuum polarization in strong
electromagnetic field. Unlike Born-Infeld electrodynamics, this theoretical model possess vacuum birefringent
properties in strong field.

Effective Lagrangian function for  Heisenberg-Euler theory  has the following form:
\begin{eqnarray}\label{HE_Lagr}
   L={I_{(2)}\over 16\pi }-{\alpha B_c^2\over
8\pi^2}\int\limits_0^\infty {e^{-\sigma} d\sigma \over \sigma^3 }
\Big[xy\sigma ^2\hbox{ctg}(x\sigma )\hbox{cth}(y\sigma)\nonumber \\
+{\sigma ^2\over 3}(x^2-y^2)-1 \Big]d\sigma,
\end{eqnarray}
where $B_c=m^2c^3/e\hbar =4.41\cdot 10^{13}\, G\,$ is the value of characteristic field in quantum
electrodynamics, $e$ and $m$ are the electron charge and mass, $\alpha=e^2/\hbar c$ -- fine structure constant
and for brevity we use the notations
\begin{eqnarray}
x=-{i\over \sqrt{2}B_c}\Big\{ \sqrt{{1\over 2}({\bf B}^2-{\bf
E}^2)+i({\bf B\ E})}-\nonumber \\
-\sqrt{{1\over 2}({\bf B}^2-{\bf E}^2)-i({\bf B\ E})}\ \Big\},
\end{eqnarray}

\begin{eqnarray}
y={1\over \sqrt{2}B_c}\Big\{ \sqrt{{1\over 2}({\bf B}^2-{\bf
E}^2)+i({\bf B\ E})}+ \nonumber \\
+\sqrt{{1\over 2}({\bf B}^2-{\bf E}^2)-i({\bf B\ E})}\ \Big\}.
\end{eqnarray}

Many attempts to find out the experimental status for each of these theories were taken for a long time, but nowadays it still
remains ambiguous. There are experimental evidences in favor of each of  them. Heisenber-Euler electrodynamics
predictions were experimentally proved in Delbr\"{u}ck light-by-light scattering \cite{a31}, nonlinear Compton
scattering \cite{a32}, Schwinger pair production in multiphoton scattering \cite{a1}. At the same time the
recent astrophysical observations \cite{a33,a34} point on the absence of vacuum birefringence effect which
favors the Born-Infeld theory prediction. The measurements performed for the speed of light in vacuum show that it doest'n depend on wave polarization with the accuracy $\delta c/c<10^{-28}$.
So clarification of vacuum nonlinear electrodynamics
status requires the expansion of the experimental test list both in terrestrial and astrophysical conditions.
The main hopes on this way are assigned to the experiments with ultra-high intensity laser facilities
\cite{a4} and astrophysical experiments with X-ray polarimetry \cite{a35} in pulsars and magnetars
neighborhood.

As it follows from Lagrangians (\ref{BI_Lagr})-(\ref{HE_Lagr}) vacuum nonlinear electrodynamics influence
becomes valuable only in~strong electromagnetic fields comparable to $E,B \sim 1/a$ for Born-Infeld theory and
$E,B \sim B_c$ for Heisenberg-Euler electrodynamics \cite{a36}.  In case of relatively weak fields ($E,B <<
B_c$) the exact expressions (\ref{BI_Lagr}) and (\ref{HE_Lagr}) can be decomposed and  written \cite{a37} in
the form of unified parametric post-Maxwellian Lagrangian:

\begin{equation}\label{PM_Lagr}
L=\frac{1}{32\pi} \Big\{2I_{(2)}+\xi\Big[(\eta_1-2\eta_2)I_{(2)}^2+ 4\eta_2I_{(4)}\Big]\Big\},
\end{equation}
where  $\xi =1/B_c^2=0.5\cdot 10^{-27} \ G^{-2}$, and the post-Maxwellian parameters $\eta_1 $ and $ \eta_2$
depend on choice of theoretical model. In case of Heisenberg-Euler electrodynamics post-Maxwellian parameters
$\eta_1 $ and $ \eta_2$ are coupled to fine structure constant $\alpha$  \cite{a38}:

\begin{equation}\label{PM_param_HE}
\eta_1={\alpha\over45\pi}=5.1\cdot 10^{-5}, \quad \eta_2={7\alpha\over180\pi}=9.0\cdot 10^{-5}.
\end{equation}

For Born-Infeld electrodynamics these parameters are equal to each other and can be expressed through the
field induction $1/a$ typical of this theory \cite{a38}:
\begin{equation}\label{PM_param_BI}
\hfil \eta_1=\eta_2={a^2 B_c^2\over 4}<4.9\cdot 10^{-6}.
\end{equation}
The electromagnetic field equations for the post-Maxwellian vacuum electrodynamics with the Lagrangian
(\ref{PM_Lagr}) are equivalent \cite{a36} to equations of Maxwell electrodynamics of continuous media
\begin{equation}\label{PM_Eq_Hom}
\partial_m F_{ik}+\partial_i F_{km}+\partial_k F_{mi}=0,
\end{equation}
\begin{equation}\label{PM_Eq_NHom}
\frac{\partial Q^{ki}}{\partial x^i}=-\frac{4\pi}{c}j^{k},
\end{equation}
with specific nonlinear constitutive relations \cite{a18}:
\begin{equation}\label{Const_Rel}
Q^{ki}=F^{ki}+\xi\Big[(\eta_1-2\eta_2) I_{(2)}F^{ki}+4\eta_2F^{ki}_{(3)}\Big],
\end{equation}
where $F^{ki}_{(3)}=F^{kn}F_{nm}F^{mi}$ is the third power of the electromagnetic field tensor. Tensor $Q^{ik}$ can be separated into two terms $Q^{ki}=F^{ki}+M^{ki}$, one of which  $M^{ki}$ will have a meaning similar to substance polarization
tensor in electrodynamics of continuous media.

Also it should be noted that in post-Maxwellian approximation stress-energy tensor $T^{ik}$ and Pointing vector
${\vec S}$ have a form:
\begin{eqnarray}\label{En-Mom_Tens}
T^{ik}={1\over4\pi}\Big\{(1+\xi\eta_1I_{(2)})F_{(2)}^{ik}-{g^{ik}\over8}
\Big[2I_{(2)}+\nonumber \\
+\xi(\eta_1+2\eta_2)I_{(2)}^2-4\eta_2\xi I_{(4)}\Big]\Big\},
\end{eqnarray}
\begin{equation}\label{Point}
S^\mu=c T^{0\mu}={c\over4\pi}\Big[1+\xi\eta_1I_{(2)}\Big]F_{(2)}^{0\mu},
\end{equation}
where $F_{(2)}^{ik}=g^{ni}F_{nm}F^{mk}$  is the second power of the electromagnetic field tensor, $g^{ik}$ --
is the metric tensor and the greek index takes a value $\mu=1,2,3$.

As it was shown in \cite{a39} that post-Maxwellian approximation turns out to be very convenient for vacuum
nonlinear electrodynamics analysis, so we will use this representation (\ref{PM_Eq_Hom})-(\ref{Point}) to
calculate radiation of the rapidly rotating pulsar.

\section{Rapidly rotating pulsar radiation in post-Maxwellian nonlinear electrodynamics}\label{Sect3}
Pulsars are the compact objects best suited for vacuum nonlinear electrodynamics tests in astrophysics. They
possess sufficiently strong magnetic field with the strength varying  from $B_p\sim10^{9}G$ up to $B_p\sim10^{14}G$, so as this values are
close to $B_c$ the vacuum nonlinear electrodynamics influence can be manifested. At the same time, the pulsar's fast rotation may enhance nonlinear influence on its radiation.

Let us consider a pulsar of radius $R_s$, rotating around an axis passing through its center with the angular
velocity $\omega$. We shall suppose that the rotation is fast enough, so the linear velocity for the points on
the pulsar's surface is comparable to speed of light $\omega R_s /c \sim 1$. We assume that pulsar's magnetic
dipole moment $\bf{m}$ is inclined to the rotation axis at the angle $\theta_0$, therefore cartesian
coordinates of this vector varies under rotation as ${\bf m}=\{m_x=m\sin\theta_0\cos\omega t, \ m_y=m\sin\theta_0\sin\omega t,
\ m_z=m \cos \theta_0\}$.

As the vacuum nonlinear electrodynamics influence in  post-Maxwellian approximation has the character of a
small correction to Maxwell theory one can represent the total electromagnetic field tensor $F^{ki}$ in form
\begin{equation}\label{Fik_Tot}
F^{ki}=F^{ki}_{(0)}+f^{ki},
\end{equation}
where $F^{ki}_{(0)}$ is the electromagnetic field tensor of the rotating magnetic dipole ${\bf m}$ in Maxwell
electrodynamics and $f^{ik}$ -- is the vacuum nonlinear correction. Substituting (\ref{Fik_Tot}) in to
(\ref{Const_Rel}) and retaining only the terms linear in small value $f^{ik}$ it can be found that:
\begin{equation}\label{Qik_lin}
Q^{ik}\simeq f^{ik}+F^{ik}_{(0)}+M^{ik}_{(0)},
\end{equation}
where $M^{ik}_{(0)}=M^{ik}(F_{(0)}^{nj})$ -- polarization tensor calculated in approximation of the Maxwell
electrodynamics field $F^{nj}_{(0)}$. Electromagnetic field equations (\ref{PM_Eq_Hom})-(\ref{PM_Eq_NHom})
with the account of (\ref{Fik_Tot})-(\ref{Qik_lin}) then will take a form:
\begin{eqnarray}\label{Total_sys}
\partial_m F_{ik}^{(0)}+\partial_i F_{km}^{(0)}+\partial_k F_{mi}^{(0)}
+ \partial_m f_{ik}+\partial_i f_{km}+ \nonumber \\
+\partial_k f_{mi}=0, \nonumber  \\
\frac{\partial f^{ki}}{\partial x^i}+\frac{\partial F^{ki}_{(0)}}{\partial x^i}+ \frac{\partial
M^{ki}_{(0)}}{\partial x^i}=-\frac{4\pi}{c}j^{k}.
\end{eqnarray}
The solution of these equation may be obtained by  successive approximation method. In initial approximation
we assume that $F_{mi}^{(0)}$ is the solution of Maxwell electrodynamics equations
\begin{eqnarray}
\partial_m F_{ik}^{(0)}+\partial_i F_{km}^{(0)}+\partial_k
F_{mi}^{(0)}=0, \nonumber \\
\frac{\partial F^{ki}_{(0)}}{\partial x^i}=-\frac{4\pi}{c}j^{k},
\end{eqnarray}
corresponding to rotating magnetic dipole ${\bf m}$, current density for which $j^k$ is represented in right
hand side of these equations. In this case, from (\ref{Total_sys}) follows that vacuum nonlinear
electrodynamics corrections $f_{ik}$ my be obtained as a solution of linearized equations:
\begin{eqnarray}
\partial_m f_{ik}+\partial_i
f_{km}+\partial_k f_{mi}=0, \label{NED_Hom} \\
\frac{\partial f^{ki}}{\partial x^i}+\frac{\partial M^{ki}_{(0)}}{\partial x^i}=0. \label{NED_NHom}
\end{eqnarray}
To satisfy homogeneous equation (\ref{NED_Hom})
electromagnetic potential $A^k$ should be introduced  $f_{ki}=\partial_k A_i-\partial_i A_k$. Using this potential the
inhomogeneous equation (\ref{NED_NHom}) under the Lorentz gauge
will take a form
\begin{equation} \label{Pot_eq}
\partial_n\partial^n A^k=\frac{\partial M^{ki}_{(0)}}{\partial x^i}.
\end{equation}
It is  more convenient to rewrite the last equation in terms of the antisymmetric Hertz tensor $\Pi^{ki}$
defined as:
\begin{equation}
A^k=-\frac{\partial \Pi^{ki}}{\partial x^i}.
\end{equation}
In this case equation (\ref{Pot_eq}) will take a simple form
\begin{equation}\label{Hertz_pot_eq}
-\partial_n\partial^n \Pi^{ki}=\Box\Pi^{ki}= M^{ki}_{(0)},
\end{equation}
where $\Box=-\partial_n\partial^n$ -- is D'Alembert operator.
Six independent equations in
(\ref{Hertz_pot_eq}) may be expressed in vector form
by introducing Hertz electric ${\bm \Pi}$ and magnetic
${\bf Z}$ potentials \cite{a40}:
\begin{equation}\label{Hertz_pot_def}
{\bm \Pi}^\alpha=\Pi^{\alpha 0},\quad {\bf Z}^\alpha=\frac{1}{2}\epsilon^{\alpha\mu\nu}\Pi_{\mu\nu},
\end{equation}
where $\epsilon^{\alpha\mu\nu}$ -- Levi-Civita symbol and all of the indexes take values
$\alpha, \mu, \nu=1,2,3$. In terms of these potentials  equations (\ref{Hertz_pot_eq}) can be rewritten:
\begin{equation}\label{Hertz_vect_eq}
\Box{\bm \Pi}={\bf P}_0,\quad \Box{\bf Z}={\bf M}_0,
\end{equation}
where the source vectors ${\bf P}_0$ and ${\bf M}_0$ are expressed from polarization tensor $M^{(0)}_{ik}$ by
equalities:
\begin{equation}\label{PM_Vec}
{\bf P}_0^\alpha=M^{\alpha 0}_{(0)} , \quad {\bf
M}_0^\alpha=\frac{1}{2}\epsilon^{\alpha\mu\nu}M^{(0)}_{\mu\nu}.
\end{equation}
The explicit components of these vectors  may be easily  obtained in Minkowski space-time with the using of
(\ref{Const_Rel}) and (\ref{PM_Vec}):
\begin{equation}\label{P0}
{\bf P}_0=2\xi \{\eta_1({\bf  E}_0^2-{\bf B}_0^2){\bf E}_0+2\eta_2 ({\bf B}_0\ {\bf E}_0){\bf B}_0\},
\end{equation}
\begin{equation}\label{M0}
 {\bf M}_0=
2\xi \{\eta_1({\bf E}_0^2-{\bf B}_0^2){\bf B}_0-2\eta_2 ({\bf B}_0\ {\bf E}_0){\bf E}_0\},
\end{equation}
where ${\bf E}_0$ and $\bf{B}_0$ are the electromagnetic field components of the rotating magnetic dipole in
Maxwell electrodynamics, the expressions for which are well described in literature \cite{a41} and the field
vectors themselves have the form:
\begin{eqnarray}\label{B0}
{\bf B}_0({\bf r}, t)={3({\bf m}(\tau)\ {\bf r}){\bf r} -r^2{\bf
m}(\tau)\over r^5}-{{ \dot{\bf m}}(\tau)\over c r^2}+ \nonumber \\
+{3({ \dot{\bf m}}(\tau)\ {\bf r}){\bf r}\over c r^4} +{(\ddot{\bf m}(\tau)\ {\bf r}){\bf r}-r^2\ddot{\bf
m}(\tau)\over c^2 r^3},
\end{eqnarray}
\begin{equation}\label{E0}
{\bf E}_0({\bf r},t)={[{\bf r},\dot{\bf m}(\tau)]\over c r^3}+ {[{\bf r},\ddot{\bf m}(\tau)]\over c^2 r^2},
\end{equation}
where $\tau=t-r/c$ is the retarded time and the dot corresponds to the derivative of the magnetic dipole
moment ${\bf m}(\tau)$ with the respect to the retarded time $\tau$. Therefore, the right hand side of the
equations (\ref{Hertz_vect_eq}) can be obtained by using of (\ref{P0})--(\ref{E0}).  The equations
(\ref{Hertz_vect_eq}) themselves  are the unhomogeneous hyperbolic equations the exact solution methods of
which are well developed and described in literature \cite{a42,a43,a44}. Since we are interested only with the
radiative solutions for the pulsar's field, when solving equations (\ref{Hertz_vect_eq})  one should retain
only the terms decreasing not faster than $\sim 1/r$ with the distance to the pulsar. At the same time there
is no restrictions on the rotational velocity so $\omega R_s/c\sim 1$. Due to excessive unwieldiness here we
will not represent the whole solutions for the Hertz potentials ${\bm \Pi}$ and ${\bf Z}$, but we will use the
results for them to find the components of the electromagnetic  field tensor $f_{ik}$ and radiation properties
such as Pointing vector ${\bf S}$ and tonal intensity $I$. The Pointing vector components represented by
(\ref{Point}) in post-Maxwellian electrodynamics, can be simplified by the radiative asymptotic condition
$S^\mu\sim 1/r^2$ which actually means that for radiation description we can use the Maxwellian expression for
this vector:
\begin{equation}\label{Point_simpl}
S^\mu=c T^{0\mu}\sim{c\over4\pi}F_{(2)}^{0\mu}.
\end{equation}
Finally, the total intensity can be obtained by integrating of Pointing vector by the surface with the normal
${\bf n}$ directed to the observer located at the large distance $r>>R_s$ from the pulsar:
\begin{equation}
I=\int ({\bf S} \  {\bf n})r^2 d\Omega,
\end{equation}
where $d\Omega$ is the solid angle.

Solutions of equations (\ref{Hertz_vect_eq}) with the right hand side (\ref{P0}),\ (\ref{M0}) lead to the
following expression for the pulsar radiation intensity:
\begin{eqnarray}\label{I_total}
I &&=\frac{2\omega^4 B_p^2 R_s^6}{3c^3}\sin^2\theta_0\Big\{1+
\frac{2}{35Y^3}\frac{B_p^2}{B_c^2}\Big( 24Y^9\Big[\frac{1}{15}\eta_1-\eta_2 \Big] \nonumber\\*
&&\times Ci(2 Y)\sin^2\theta_0 +
4Y^9\Big[ \eta_2-\frac{311}{45}\eta_1 \Big]Ci(2 Y)  \nonumber\\*
&&+Y^3\Big[\frac{\eta_1-15\eta_2}{5} (2Y^4-3Y^2)
- 18\eta_1 -10\eta_2\Big]\cos(2Y) \sin^2 \theta_0 \nonumber\\*
&&+\frac{Y}{6}\Big[\frac{45\eta_2-311\eta_1}{15}
(2Y^6-3Y^4)+(172\eta_1- 60\eta_2)Y^2 \nonumber\\*
&&- 336\eta_1 \Big]\cos(2Y)
+Y^2\Big[ \frac{30\eta_2 -2\eta_1}{5} (2Y^6-Y^4)\nonumber \\
&&-\frac{41\eta_1+85\eta_2 }{5}Y^2+ 9\eta_1 + 5\eta_2 \Big]\sin(2 Y)
\sin^2\theta_0\nonumber\\
&&+ \frac{1}{3}\Big[\frac{311\eta_1-45\eta_2}{15}(2Y^8-Y^6 +3Y^4)\nonumber\\
&&+(15\eta_2- 141\eta_1) Y^2 + 84\eta_1\Big]\sin(2 Y)\Big)\Big\},
\end{eqnarray}
where the following notations are used for brevity:  $k=\omega/c$ and $Y=k R_s$, also $B_p$ -- is the surface
magnetic field inductance and $Ci(x)=\int\limits_\infty^x\frac{\cos u}{u} du $ -- is an integral cosine.

It is obvious that obtained intensity can be represented in form which distinguishes Maxwell radiation
intensity and vacuum nonlinear electrodynamics correction. In this representation it is convenient to
introduce the "correction function" $\ \Phi(\theta_0,Y)$ which is a multiplier before the scaling factor
${B_p^2}/{B_c^2}$ determining how strong the vacuum nonlinear electrodynamics influence on the pulsar
radiation is:
\begin{equation}
I=\frac{2\omega^4 B_p^2 R_s^6}{3c^3} \sin^2\theta_0\Big\{1+ \frac{B_p^2}{B_c^2}\ \Phi(\theta_0,Y)\Big\}.
\end{equation}
For the most known rapidly rotating pulsars \cite{a45} with $Y\sim 1$ the factor ${B_p^2}/{B_c^2}\ll 1$ is
small which matches the requirements of post-Maxwellian approximation. At the same time this means that vacuum
nonlinear electrodynamics corrections will be sufficiently suppressed in comparison with Max\-well
electrodynamics radiation. However, this assessment may be waived for special sources of so called Fast Radio
Bursts (FRB's), six cases of which have recently been discovered \cite{a46}. One of the hypotheses explaining
the nature of FRBs assumes that their source is a rapidly rotating neutron star with the strong surface
magnetic field $B_p>B_c$ called blitzar \cite{a55}.
In this case vacuum nonlinear electrodynamics corrections to pulsar
radiation became significant but at the same time this makes strict solution (\ref{I_total}) unapplicable
because it was obtained in low-field limit. So our further evaluations will be applied to the case of
the typical rapidly rotating pulsar, for instance PSR B1937+21 with $B_p\sim 4.2 \cdot 10^8G \ll B_c$, and
maybe for blitzars but with the restriction $B_p<B_c$. The main purpose of our analysis will be in identification of
new qualitative features of the pulsar radiation  and comparing vacuum nonlinear corrections to the
electromagnetic radiation with the other weak energy loose mechanisms.

\begin{figure}
\includegraphics[width=0.5\textwidth]{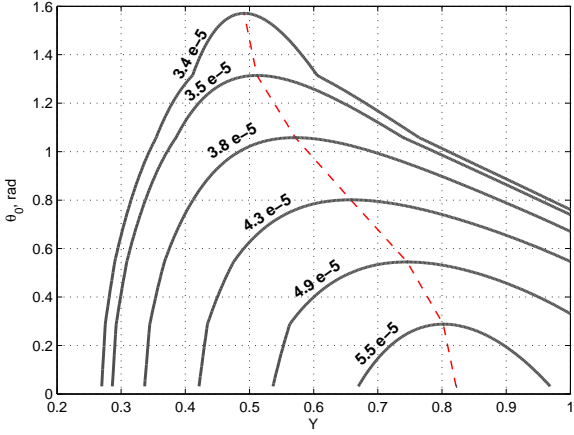}
\caption{Correction function isolines and the best contrast line}
\label{ContrFig}       
\end{figure}

Let's investigate the properties of the correction function $\Phi(\theta_0,Y)$.
First of all, it should be noted that there is no radiation when the pulsar dipole moment is coaxial with the
rotation axis i.e. when $\theta_0$ is zero. The correction function depends both on the angle $\theta_0$ and
the angular velocity through $Y=\omega R_s/c$, so the $\Phi(\theta_0,Y)$ may be represented as a surface
defined in the region where it's coordinates take values $0\leq Y<1$ and $0\leq\theta_0\leq \pi/2$. Some
isolines -- the relations $\theta_0(Y)$ at which this surface takes a constant value $\Phi(\theta_0,Y)=const$
are represented at the Fig.\ref{ContrFig},
the numerical values for which
were obtained  with the $\eta_1$ and $\eta_2$
from the Heisenberg-Euler theory.

The obtained isolines differ from each other by the absolute value of the correction function but all of them
have pronounced extremum at some point which lies on the red line.
This means that for each fixed angle
$\theta_0$ between the pulsar dipole moment and rotation axis there is an angular velocity at which vacuum
nonlinear electrodynamics corrections  become the most pronounced. Increasing $Y$ at the constant $\theta_0$
up to the value marked by the red line increases correction of vacuum nonlinear electrodynamics. The
subsequent $Y$ and angular velocity increase becomes ineffective because the vacuum corrections in this case
will be reduced. It should be noted, that increasing $Y\to 1$ also will enhance total pulsar luminosity which is
$I\propto \omega^4 \sin^2\theta_0$ but at the same time, as it was mentioned, this will decrease vacuum
nonlinear electrodynamics correction on the Maxwell radiation background. For instance, if $\theta_0\sim
\pi/2$ the correction will  most significantly stand out for the pulsars with $Y\sim 0.5$. So the correction
function $\Phi(\theta_0,Y)$ plays a role of a contrast. And the red line in Fig.\ref{ContrFig} marks the
relation between $\theta_0$ and $Y$ for the best contrast.

Another distinctive feature of the pulsar radiation is manifested in    sophisticated, non-polynomial dependence
between the radiation intensity (\ref{I_total}) and the angular velocity, which greatly complicates the
analysis. Performing power-law approximation of (\ref{I_total}) will allow us to describe vacuum nonlinear
electrodynamics influence on the pulsar spin-down, in traditional terms of braking-indexes and
torque-functions \cite{a47}. It also provides a possibility for comparison of the pulsar spin-down caused by
different non-electromagnetic dissipative factors and the power-low relation between the radiation intensity
and angular velocity, for instance with the quadrupole gravitational radiation. Let us investigate the features
of the pulsar spin-down as a result of the radiation, with the amendments of vacuum nonlinear electrodynamics.

\section{Pulsar spin-down}\label{Sect4}
The observed spin-down rate  \cite{a47} can be expressed by derivative of the angular velocity as
\begin{equation}\label{SpD_initial}
\dot{\omega}=-\frac{I}{J\omega}=-\frac{2B_p^2R_s^3}{3J}\sin^2\theta_0
\Big\{\,Y^3+\frac{B_p^2}{B_c^2}\,Y^3\,\Phi \Big\},
\end{equation}
where $J$ is the pulsar's inertia momentum and the dot means the time derivative. For description in terms of
torque functions, the right hand side of  equation (\ref{SpD_initial}) should be represented in polynomial
form of angular velocity
\begin{equation}\label{Decomp}
Y^3\Phi(\theta_0, Y)=\sum\limits_n^N\alpha_n(\theta_0)\,Y^n=
\sum\limits_n^N\alpha_n(\theta_0)\,\Big(\frac{\omega R_s}{c}\Big)^n,
\end{equation}
where $\alpha_n$ are decomposition coefficients and  the number of the terms $N$ should be selected sufficient
to ensure the required accuracy of the decomposition. We will take the  number of terms in the expansion
(\ref{Decomp}) equal to $N=8$. This  choice ensures the accuracy of power-law approximation for the  pulsars
with the  $Y>0.6$ better then 0.1\%. It should be noted, that the series does not converge at  $Y\sim1$ but it's replacement by the partial sum with the specially selected number of terms allows to accomplish the polynomial approximation which provides a good match with the exact expression near $ Y\sim 1 $, but leads to significant errors when $Y\ll1$.
In this case the expansion coefficients (with the $\eta_1$ and $\eta_2$ from the Heisenberg-Euler theory) for the terms providing the largest contribution are represented in the Table \ref{tab:1}.
The coefficients  not listed in the table are small and can
be neglected in further consideration.
\begin{table}
\caption{Expansion coefficients}\label{tab:1}
\begin{tabular}{l|lllll}
\hline\noalign{\smallskip} $\theta_0$ &$\alpha_7\cdot 10^{4}$ & $\alpha_6\cdot 10^{4}$ &  $\alpha_5\cdot
10^{4}$
& $\alpha_4 \cdot 10^{4}$ & $\alpha_3 \cdot 10^{4}$\\
\noalign{\smallskip}\hline\noalign{\smallskip}
$\pi/2$& $1.7  $ & $-4.0$ & $2.4$ & $-0.3 $& $-0.2$\\
$\pi/3$& $1.5  $ & $-3.8$ & $2.7$ & $-0.5 $& $-0.3$\\
$\pi/6$& $1.1  $ & $-3.6$ & $3.3$ & $-0.9 $& $-0.5$\\
\noalign{\smallskip}\hline
\end{tabular}
\end{table}
For quantitative analysis, we will take the inclination angle is equal to $\theta_0=\pi/2$. This choice is
justified because it provides the greatest total intensity of the pulsar radiation and in our comparison of
nonlinear electrodynamics spin-down with the other non-electrodynamics dissipative factors, it gives the upper
limit of nonlinear electrodynamics influence.

After the expansion, the right hand side of the spin-down equation (\ref{SpD_initial}) will take the form:
\begin{equation}\label{SpD_torque}
\dot{\omega}=K_M+\sum\limits_n K_n\omega^n,
\end{equation}
where $K_M$ corresponds to the torque function of the dipole magnetic radiation in Maxwell electrodynamics
\cite{a48,a49}:
\begin{equation}\label{KM}
K_M=-\frac{2B_p^2R_s^6}{3Jc^3}\sin^2\theta_0,
\end{equation}
and $K_n$ are the torques originating form nonlinear vacuum electrodynamics:
\begin{equation}\label{Kn}
K_n=\alpha_n(\theta_0) K_M\Big(\frac{B_p}{B_c}\Big)^2\Big(\frac{R_s}{c}\Big)^{n-3}.
\end{equation}

Let us compare pulsar spin-down caused by nonlinear vacuum electrodynamics and dissipation caused by
gravitational waves radiation. Among several possible ways of  gravitational radiation by an isolated pulsar
we will choose two most relevant scenario -- quadrupole mass radiation \cite{a50} and the radiation caused by
Rossby waves \cite{a51}, called r-modes.

Quadrupole gravitational radiation can be originated by the strain caused by the pulsar rotation, which is
especially likely for rapidly rotating pulsars. The spin-down under this kind of radiation can be represented
as:
\begin{equation}\label{GW_quadr}
\dot{\omega}=K_{Q}\;\omega^5=-{32\over 5}{GJ\varepsilon^2 \over c^5}\;\omega^5,
\end{equation}
where $G$ is a gravitational constant and $\varepsilon$ is the pulsar elipticity, which is in accordance with
modern representations $\varepsilon<10^{-4}$ \cite{a47}.

Another reason for gravitational waves emission by isolated pulsar are the oscillations modes induced by the
pulsar rotation. Gravitational radiation is caused by instability of such oscillations. As it was shown by
Owen et al. \cite{a52} for young rapidly rotation pulsars spin-down  caused by r-modes can be expressed in
form:
\begin{equation}\label{GW_r-modes}
\dot{\omega}=K_{R}\;\omega^7=-\frac{ 2^{17} \pi F^2 GM^2R_s^6 \beta_{sat}^2} {3^7  5^2  J c^7}\;\omega^7,
\end{equation}
where $M$ and $R_s$ are the pulsar mass and it's radius, the r-mode oscillations saturation amplitude
$10^{-7}\leq\beta_{sat}\leq 10^{-5}$ was defined by \cite{a53}, and dimensionless constant $F$ as it has been
shown in \cite{a54} is to be strictly bounded within $1/(20\pi)\leq F\leq 3/(28 \pi)$.

So the torque function for the quadrupole gravitational radiation $K_{Q}$ can be compared with  nonlinear
electrodynamics torque $K_5$ and the r-modes radiation torque $K_{R}$ can be compared with the torque $K_7$.
For this comparison we suppose the pulsar with the typical radius $R_s=30\ km$, mass $M=2M_\odot$ and inertia
momentum $J=10^{45}\  g\cdot cm^2$. Also we assume, that dipole moment inclination is $\theta_0=\pi/2$ and
post-Maxwellian parameters correspond to Heisenberg-Euler theory (choice of Born-Infeld parameters in first
estimation gives the similar order).

For the pulsar with the surface magnetic field $B_p\sim 10^{11} G$, which elipticity reaches the maximum value
$\varepsilon\sim 10^{-4}$, r-mode saturation amplitude $\beta_{sat}\sim 10^{-6}$, and $F=1/(20\pi)$ the
following estimation takes place  $K_5/K_Q\sim 1.3 \cdot 10^{-11}$ and $K_7/K_R \sim 2.6 \cdot 10^{-7}$. So
the quadrupole and r-mode gravitational radiation torque will significantly exceed nonlinear electrodynamics
torque coupled with the terms $\sim \omega^5$ and $\sim \omega^7$ in spin-down equation. For another parameter
set the opposite case takes place. If the pulsar distortion and elipticity is two orders of magnitude lower (
$\varepsilon\sim 10^{-6}$), and the pulsar field is stronger $B_p\sim 10^{13}G$ then $K_5/K_Q\sim 12.6$ and
$K_7/K_R\sim 25.6$. However, it should be noted that the rapidly rotating pulsars with such a strong field
have not been observed yet. Nevertheless, the theoretical models assuming the  blitzars  as the sources of Fast
Radio Burst \cite{a55} do not eliminate the possibility of  such a  strong electromagnetic fields for the
rapidly rotating pulsar. Therefore obtained  ratio between the torques seems very exotic but still can not be
completely discarded.

\section{Conclusion}\label{Sect5}
In this work, we have studied vacuum nonlinear electrodynamics
influence on rapidly rotating pulsar radiation
in parameterized post-Maxwellian electrodynamics. In assumption of
flat space-time the analytical description of
radiation intensity (\ref{I_total}) was obtained. Despite on the fact
that the expression for the intensity is
quite complicating for analysis some new features of pulsar's radiation
have been obtained. For instance,  it
was shown that for the rapidly rotating pulsar vacuum nonlinear
electrodynamics corrections observation is
optimal only for certain relations between the inclination angle $\theta_0$ of the magnetic dipole moment to the
rotation axis and the angular velocity $\omega$.
Such relation plays a role of  the contrast for
nonlinear corrections  on  total  pulsar radiation background.
It follows that enhancing of vacuum nonlinear
electrodynamics influence on pulsar radiation requires not only
increasing magnetic field but also needs
compliance of conditions marked on Fig.\ref{ContrFig}
to ensure the best possible contrast
for the nonlinear corrections.

The obtained  radiation intensity was used to estimate
pulsar spin-down. In this framework, for description in
terms of the torque functions the power-low expansion
of the intensity (\ref{I_total}) was carried out
(\ref{SpD_torque})-(\ref{Kn}) with the decomposition
coefficients listed in Table \ref{tab:1}. This provided
an opportunity to compare nonlinear electrodynamics
torque with the the weak mechanisms of the energy
dissipation, for instance with  gravitational waves radiation.
For such comparison the most realistic
scenarios of  gravitational  radiation  by isolated pulsar
were selected  -- quadrupole gravitational radiation
and r-modes radiation. The quantitative comparison has
shown that for the  common rapidly rotating pulsar,
gravitational radiation torques significantly exceed
nonlinear  electrodynamics torques coupled with the terms
of same $\omega$ powers in spin-down equation. This
result can be explained by low surface magnetic
field $B_s<10^{11} G$ specific for most of rapidly rotating
pulsar's population. Implementation of similar estimates
for the compact object possessing stronger magnetic
field (hypothetical blitzar) $B_s\sim 10^{13} G$ shows the
possibility of the opposite case when vacuum nonlinear
electrodynamics torques exceed gravitational torque and
play more significant role in spin-down equation under certain conditions.


\begin{thebibliography}{99}
\bibitem{a1} D. L. Burke et al. Phys. Rev. Lett. {\bf 79}, 1626 (1997)
\bibitem{a2} V. I. Denisov, I. P. Denisova, S. I. Svertilov,
 Theoretical and Mathematical Physics {\bf 135}, 720 (2003)
\bibitem{a3} G. O. Schellstede, V. Perlick,  C. L$\ddot{\hbox{a}}$mmerzahl,
Phys. Rev. D, {\bf 92}, 025039 (2015)
\bibitem{a4} F. D. Valle et. al., Eur. Phys. J. C, {\bf 76}, 24 (2016)
\bibitem{a5} G.V. Dunne, Eur. Phys. J. D {\bf 55}, 327 (2009).
\bibitem{a6}  \url{http://www.eli-beams.eu/}
\bibitem{a7} G. Mourou, T. Tajima, Optics $\&$ Photonics News {\bf 22}, 47 (2011)
\bibitem{a8}  \url{http://www.hzdr.de/db/Cms?pOid=35325&pNid=3214}
\bibitem{a9} V. I. Denisov, I. P. Denisova, Optics and Spectroscopy {\bf 90}, 928 (2001)
\bibitem{a10} A. Paredes, D. Novoa, D. Tommasini, Phys. Rev. Lett. {\bf 109}, 253903 (2012)
\bibitem{a11} V. I. Denisov, I. P. Denisova, Theoretical
and Mathematical Physics, {\bf 129}, 1421 (2001)
\bibitem{a12} J. Schwinger, Phys. Rev., {\bf 82}, 664 (1951)
\bibitem{a13} S.L. Adler, Ann. Phys., {\bf 67}, 599 (1971)
\bibitem{a14} P. A. Vshivtseva, V. I. Denisov, I. P. Denisova, P. Doklady Physics, {\bf 47}, 798 (2002)
\bibitem{a15} V. B. Berestetskii, L. P. Pitaevskii,
E.M. Lifshitz, Quantum Electrodynamics,
(Pergamon Press, Oxford, UK, 1982)
\bibitem{a16} V. I. Denisov, I. P. Denisova, S.I.Svertilov,
Doklady Physics, {\bf 46}, 705 (2001)
\bibitem{a17} J.Y. Kim,  Journal of Cosmology and Astroparticle Physics, {\bf 10}, 056 (2011)
\bibitem{a18} V. I. Denisov, Theoretical and Mathematical Physics, {\bf 132}, 1071, (2002)
\bibitem{a19} M. G. Baring, A. K. Harding,  ApJ, {\bf 547}, 929 (2001)
\bibitem{a21} M. Born, L. Infeld , Proc.  Roy.  Soc.,  {\bf A144}, 425 (1934)
\bibitem{a22} M. Born, L. Infeld , Proc.  Roy.  Soc.,  {\bf A147}, 522 (1934)
\bibitem{a23} J. B. Kogut, D. K. Sinclair, Phys. Rev. D {\bf 73}, 114508, (2006)
\bibitem{a24} E. S. Fradkin, A. A. Tseytlin, Phys.  Lett. B {\bf 163}, 123 (1985)
\bibitem{a25}  S. Cecotti, S.Ferrara, Phys. Lett. B, {\bf 187}, 335 (1987)
\bibitem{a26} B. Zwiebach, A First Course in String Theory, (Cambridge University Press., 2004)
\bibitem{a27} P. Gaete, J. Helay$\ddot{\hbox{e}}$-Neto, Eur. Phys. J. C, {\bf 74}, 3182 (2014)
\bibitem{a28} Z. Bilanicka, I. Bialynicki-Birula, Phys. Rev. D {\bf 2}, 2341 (1970)
\bibitem{a29} B. Hoffmann, L. Infeld, Phys. Rev. {\bf 51}, 765, (1937)
\bibitem{a30} W. Heisenberg,  H. Euler, Z. Phys., {\bf 26}, 714 (1936)
\bibitem{a31} R. R. Wilson, Phys. Rev. {\bf 90}, 720, (1953)
\bibitem{a32} C. Bula et al. Phys. Rev. Lett. {\bf 76}, 3116 (1996)
\bibitem{a33} Wei-Tou Ni, Physics Letters A {\bf 379}, 1297 (2015)
\bibitem{a34} Wei-Tou Ni, Hsien-Hao Mei, Shan-Jyun Wu, Modern Physics Letters A {\bf 28}, 1340013 (2013)
\bibitem{a35} P. Soffitta et al., Exp. Astron. {\bf 36}, 523, (2013)
\bibitem{a36} V.R. Khalilov, Electrons in Strong Electromagnetic Fields: An Advanced Classical
and Quantum Treatment, (Gordon and Breach Science Pub New York, Netherlands, 1996)
\bibitem{a37} V. I. Denisov, I. P. Denisova, Doklady Physics, {\bf 46},  377, (2001)
\bibitem{a38} V. I. Denisov, I. P. Denisova, Optics and Spectroscopy {\bf 90}, 282 (2001)
\bibitem{a39} V.I. Denisov, V.A. Sokolov, M.I. Vasili'ev, Phys. Rev. D, {\bf 90}, (2014)
\bibitem{a40} I. P. Denisova, M. Dalal, J. Math. Phys., {\bf 38}, 5820 (1997)
\bibitem{a41} V.I. Denisov, I. P. Denisova, V.A. Sokolov,
Theoretical and Mathematical Physics, {\bf 172}, 1321 (2012)
\bibitem{a42} J. Mathews,R.L. Walker,  Mathematical methods of physics (2nd ed), New York: W. A. Benjamin, 1970)
\bibitem{a43} K.V. Zhukovski, Moscow University Physics Bulletin, {\bf 70}, 93, (2015)
\bibitem{a44} K.V. Zhukovski, Moscow University Physics Bulletin, {\bf 71}, 237, (2016)
\bibitem{a45} R.N. Manchester et al., ATNF pulsar catalog (Manchester, 2005)
\bibitem{a55} H. Falcke, L. Rezzolla, Astronomy $\&$ Astrophysics, {\bf 562}, A137 (2014)
\bibitem{a46} A. Loeb, Y. Shvartzvald,  D. Maoz, MNRASL {\bf 439}, L46 (2014)
\bibitem{a47} C. Palomba, Astron. Astrophys., {\bf 354}, 163 (2000)
\bibitem{a48} J.P. Ostriker, J.E. Gunn, ApJ, {\bf 157}, 1395 (1969)
\bibitem{a49} R.N. Manchester, J.H. Taylor,  Pulsars, (San Francisco: W. H. Freeman, 1977)
\bibitem{a50}  L.D. Landau; E.M. Lifshitz,. The Classical Theory of Fields. Vol. 2 (4th ed.),(Butterworth-Heinemann,1975)
\bibitem{a51} N. Stergioulas, J. A.Font, Phys. Rev. Lett., {\bf 86}, 1148, (2001)
\bibitem{a52} B. J. Owen et al, Phys. Rev. D {\bf 58}, 084020 (1998)
\bibitem{a53} Alford M. G. and  Schwenzer K., MNRAS 446, 3631-3641, 2015.
\bibitem{a54} M. G. Alford, K. Schwenzer, ApJ, {\bf 26}, 781, (2014)
\end{thebibliography}
\end{document}